\documentclass{jpsj-suppl}
\usepackage{txfonts} %Please comment out this line unless the txfonts package is availabe in your LaTeX system.

\title{Search for a correlation between giant radio pulses and hard X-ray emissions in the Crab pulsar}

\author{Ryo \textsc{Mikami}$^{1}$, Toshio \textsc{Terasawa}$^{1;2}$, Shota \textsc{Kisaka}$^{1}$, Hideaki \textsc{Miyamoto}$^{1}$, Katsuaki \textsc{Asano}$^{1}$, Nobuyuki \textsc{Kawai}$^{2}$, Yosuke \textsc{Yamakoshi}$^{2}$, Kumiko \textsc{Nagata}$^{2}$, Ryuho \textsc{Kataoka}$^{3}$, Kazuhiro \textsc{Takefuji}$^{4}$, Mamoru \textsc{Sekido}$^{4}$, Hiroshi \textsc{Takeuchi}$^{5}$, Hirokazu \textsc{Odaka}$^{5}$, Tamotsu \textsc{Sato}$^{5}$,
and Yasuyuki T. \textsc{Tanaka}$^{6}$}

\inst{$^{1}$Institute for Cosmic Ray Research, The University of Tokyo, Chiba 277-8582, Japan \\
$^{2}$Department of Physics, Tokyo Institute of Technology, Tokyo 152-8551, Japan \\
$^{3}$National Institute of Polar Research, Tokyo 190-8518, Japan \\ 
$^{4}$Kashima Space Research Center, National Institute of Information and Communications Technology, Ibaraki 314-8501, Japan \\
$^{5}$Institute of Space and Astronautical Science, Japan Aerospace Exploration Agency, Kanagawa, 229-8510, Japan \\
$^{6}$Hiroshima Astrophysical Science Center, Hiroshima University, Hiroshima 739-8526, Japan}

\email{mikami@icrr.u-tokyo.ac.jp}

\recdate{July 15, 2013}

\abst{We present the results of the search for a correlation between giant radio pulses (GRPs) at 1.4 GHz 
and hard X-rays at 15-75 keV from the Crab pulsar. 
We made simultaneous ground and satellite observations of the Crab pulsar over 12 hours in three occasions 
in April 2010, March and September 2011, and got a sample of 1.3$\times10^{4}$ main-pulse phase GRPs. 
From these samples we have found statistically marginal enhancement (21.5\%, 2.70 $\sigma$) of hard X-ray flux 
within $\pm$ 1.5 degree phase angle of the  synchronous peak of main-pulse phase GRPs. 
This enhancement, if confirmed, implicates that GRPs may accompany plasma density increases in the pulsar magnetosphere. }

\kword{Pulsars: Crab pulsar (PSR B0531+21), Giant radio pulses, X-rays}

\begin{document}
\maketitle

\section{Introduction}
Giant radio pulses (GRPs) are a special, distinctive class of pulsar radio emission. 
The duration of GRPs is short (from nanosecond to millisecond), and 
their brightness temperatures sometimes exceed $10^{41}$ K\cite{HE07}. %The intensity distribution of GRPs is expressed by the power law.
While more than 2000 pulsars have been found so far, 
only less than 1\% of these pulsars are known to emit GRPs
\cite{RJ01, JR03, Si12}. 

The Crab pulsar, PSR B0531+21, is the most studied pulsar of those which emit GRPs. 
GRPs from the Crab pulsar have been observed from 20 MHz\cite{El13} to 15.1 GHz\cite{Je10}. 
They occur at the two phases of normal radio pulses, namely the main-pulse (hereafter MP) phase and the interpulse (hereafter IP) phase \cite{Co04}. 

\begin{table}[t]
\centering
\caption{Correlation studies between GRPs and pulses at other frequencies
from the Crab pulsar.}
\label{t1}
\begin{tabular}{c|c|c|c}
Energy range & Satellite or Telescope & Flux variation synchronous with GRPs & Ref.\\
\hline \hline
Optical  & William Herschel  & 3\% (7.8 $\sigma$) increase & \cite{Sh03}\\ 
(600-750 nm) &Telescope & & \\ \hline
 Soft X-ray& Chandra HRC-S & $<$10\% (2 $\sigma$)& \cite{Bi12}\\
(1.5-4.5 keV)& &(MP phase window) & \\ \hline
Hard X-ray  & Suzaku HXD & 21.5\% (2.70 $\sigma$) increase & This work.\\
(15-75 keV) & & & \\ \hline
Soft $\gamma$-ray & CGRO OSSE & $<$250\% (1 $\sigma$)& \cite{Lu95}\\
(50-220 keV) & & (GRP $\pm$ 5 periods)& \\ \hline
$\gamma$-ray & Fermi LAT & $<$400\% (95\% CL)& \cite{Bi11}\\
(0.1-5 GeV) & & (around GRPs) & \\ \hline
Very High Energy $\gamma$-ray & VERITAS & $<$500-1000\% (95\% CL) & \cite{Al12}\\
($>$150 GeV) & & (around IP GRPs) & \\ \hline
\hline
\end{tabular}
\end{table}%

In wide frequency ranges the Crab pulsar emits pulses.
Among them, coherently emitted pulses at radio frequencies (both normal and giant),
had been thought to be independent from 
those at higher frequencies (infrared, optical, X-ray, and gamma-ray), 
since the latter are originated in incoherent processes. 
However, Shearer et al. \cite{Sh03} discovered that
optical pulses from the Crab pulsar 
show the significant 3\% (7.8 $\sigma$) increases synchronously with MP GRPs, proving
that there is some interplay between the emission mechanisms for
pulses at radio and other frequencies.
Table \ref{t1} shows the summary of correlation studies between GRP and 
pulses at other frequencies from the Crab pulsar. 
Except for the optical study stated above\cite{Sh03}, 
the previous studies \cite{Bi12,Lu95,Bi11,Al12} set only upper limits for the enhancement 
synchronous with GRPs.
The hard X-ray energy range (the third line in Table \ref{t1}), 
where no correlation study based on simultaneous observations has been reported
to our knowledge, is a target of our study.
We obtain statistically marginal enhancement (21.5\%, 2.70$\sigma$) 
as described in the following sections.

\section{Observations}
Simultaneous observations of the Crab pulsar both at radio (1.4 GHz) and X-ray frequencies were made on 2010 April 6, 2011 March 22 and 2011 September 1-2 (Table \ref{obs}).
\begin{table}[b]
\centering
\caption{Summary of the radio and hard X-ray observations. The values of the dispersion measure, DM, used in the dedispersion procedure for the radio signals are tabulated in the third column.}
\label{obs}
\begin{tabular}{c|c|c|c|c|c}
Date & radio freq.&DM&simultaneous & No. of MP GRPs & No. of X-ray photons  \\
 &(MHz) &(cm$^{-3}$pc)&obs. time (min)& (Ratio to No. of all pulses) & \\ \hline \hline
2010 Apr. 6  & 1405-1435 &56.835 & 313 & 4090 (0.73\%)& 571763 \\ 
2011 Mar. 22 & 1400-1450$^a$ &56.800 & 178 & 2568 (0.81\%)& 323725 \\
2011 Sep. 1-2 & 1405-1435 & 56.800& 271 & 6487 (1.34\%)& 488934 \\ \hline
Total &  & & 762 & 13145 (0.97\%) & 1384422 \\ \hline
\hline
\end{tabular}
{\footnotesize{$^a$The frequency range 1426-1436 MHz is omitted for the RFI rejection.}}
\end{table}%
The radio data were acquired by the 34-m Kashima radio telescope (in April 2010 and September 2011) and by the Usuda 64-m radio telescope (in March 2011). Following the standard procedure for the radio pulsar observation \cite{LoKr05}, we dedispersed the radio signals, removed radio frequency interferences (RFIs), and then identified GRPs at MP phases above the signal-to-noise ratio of 5, where the noise level is dominated by the nebula radio emissions of $\sim$ 1 kJy. The X-ray data were acquired by the hard X-ray detector (HXD) aboard the Suzaku satellite \cite{Ta07}. In what follows, we focused on the correlation study of MP GRPs with the X-ray data (15-75keV).  

%With the data mentioned above, we compared flux densities of hard X-ray pulses coincident with MPGRPs to those not coincident with MPGRPs.
\section{Result}
\begin{figure}[tbh]
\centering
\includegraphics[width=15cm]{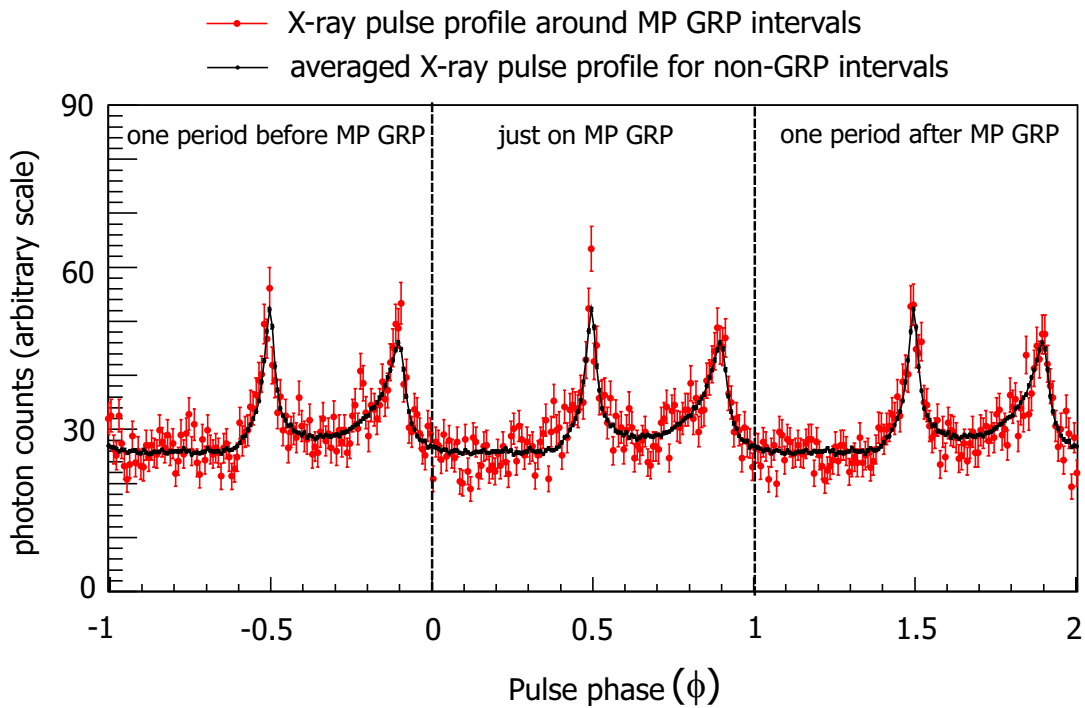}
\caption{Comparison of the hard X-ray pulse profile, in a vicinity of MP GRPs ($\phi$ = 0.5, red) and the averaged X-ray pulse profile (repeated for three successive intervals, black) (See text.). Error bars represent 1 $\sigma$ statistical uncertainties.}
\label{f1}
\end{figure}

Figure \ref{f1} shows the results of superposed epoch analyses, 
where the red points shown with statistical errors represent 
the accumulated hard X-ray photon counts over the 3 spin periods, 
for which the timing of the MP GRPs is set at the spin phase $\phi$= 0.5. 
The spin phase of 
$-1 \leq \phi < 0.5$
$ (0.5 < \phi \leq 2)$ corresponds to the 1.5 pulse intervals preceding (following) the MP GRP detections. 
The photon counts were accumulated in each of 1/120 period phase bins (namely, bins with a 3-degree phase angle width). 
In Figure \ref{f1} we repeat three times the averaged hard X-ray pulse profile obtained with normal (namely, non-GRP) radio pulses (a black curve) so as to facilitate the search for the difference, if any, between X-ray pulse profiles during GRP and non-GRP intervals. 
It is seen that the X-ray profile has statistically marginal enhancement (21.5\%, 2.70 $\sigma$) 
above the average profile in one bin at phase $\phi$ = 0.5.  
By varying the size of bins, we investigate how the significance of the enhancement depends on the choice of the bin size:
While the enhancement tends to be smeared out as the bin size is widened ($>$3 degrees),
the X-ray photon numbers become too few and their fluctuations grow as the bin size is narrowed ($<$3 degrees).
Thus observing the reduction of the statistical significance 
on the both sides of the bin size, larger or smaller than 3 degrees,
we tentatively conclude that the statistically `best' choice of the bin size is 3 degrees
for the present X-ray data sets where photon numbers are limited.

%\begin{figure}[tbh]
%\centering
%%\includegraphics[height=7cm]{spectrumAPPCver20130709.eps}
%\caption{Flux increase concurrent with GRPs in several energy bands. The green point is the data of Shearer et al. \cite{Sh03}. The blue point is the data estimated from Bilous et al. \cite{Bi12}. The red point is our data. Error bars represent 3 $\sigma$ statistical uncertainties.}
%\label{f2}
%\end{figure}%
\section{Discussion and Conclusion}
We have made a correlation study between 15-75 keV hard X-rays and main-pulse phase (MP) GRPs at 1.4GHz from the Crab pulsar. 
We have found marginal (21.5\%, 2.70 $\sigma$) enhancement of the hard X-ray flux synchronously with MP GRPs at their peak.
In the previous correlation studies, only Shearer et al. \cite{Sh03} showed a correlation between GRPs and other energy emissions, who argued that GRPs and the increased optical emission are linked to an increase in the electron-positron plasma density. Our result, if confirmed, gives a conclusion parallel to the Shearer's that the GRPs and the enhanced hard
X-ray emission are linked to the plasma density increase in the Crab pulsar magnetosphere.

In the current analysis we see the possible hard X-ray flux enhancement
 only in one bin with the width of 1/120 period (or 3 degree) at the MP GRP peak. 
We might be able to speculate that the interplay between emission processes of
GRPs and X-ray pulses occurs in some limited region of the magnetosphere.
However, it is premature to discuss this point further,
and we should wait for further simultaneous observations 
at radio and X-ray frequencies to improve statistics.

\end{document}